\documentclass{pazheng}

\usepackage{amsmath}
\usepackage{amsfonts}
\usepackage{amssymb}
\usepackage{amstext}
\usepackage{natbib}
\usepackage{indentfirst}
\usepackage{misccorr}

\bibpunct{(}{)}{;}{a}{,}{,}

\usepackage{lscape}
\usepackage{mathtext}
\usepackage{graphicx}
\usepackage{float}

\usepackage{textcomp}
\usepackage{mathtext}

\usepackage{epsfig}
\usepackage{rotating}
\usepackage{dcolumn}
\usepackage{tabularx}

\newcommand{\Msun}{M_\odot}
\newcommand{\Ho}{HoIX X-1}
\newcommand{\M}{M82 X-1}
\newcommand{\NH}{N_{\rm H}}

\begin{document}

\title{Cutoff in the hard X-ray spectra of the ultraluminous X-ray
  sources \Ho\ and \M}

\author{\bf 
S.~Sazonov \email{sazonov@iki.rssi.ru}\address{1},
A.~Lutovinov \address{1},
R.~Krivonos \address{1,2}
\addresstext{1}{Space Research Institute, Russian Academy of Sciences,
  Moscow} 
\addresstext{2}{Space Sciences Laboratory, University of California,
  Berkeley, USA}  
}
\shortauthor{}
\shorttitle{}
\submitted{\today}

\begin{abstract}
Using data from the XMM-Newton and INTEGRAL observatories, we
have detected a cutoff at energies above 10~keV in the X-ray
spectra of the ultraluminous X-ray sources \Ho\ and \M. The spectra
obtained can be described by a model of Comptonization of radiation in
a gas cloud of moderate temperature ($kT\sim 2$--3~keV) and high
optical depth ($\tau\sim 15$--25). Such conditions can be fulfilled 
during supercritical accretion of matter onto a stellar-mass black
hole accompanied by a strong gas outflow. The results of this work  
confirm the existence of a spectral state specific to ultraluminous
X-ray sources, which is unlike any of the known spectral states in
normal X-ray binaries.

\englishkeywords{black holes, accretion, Comptonization, ultraluminous X-ray
  sources}

\end{abstract}    

%%%%%%%%%%%%%%%%%%
\section{Introduction}
\label{s:intro}
%%%%%%%%%%%%%%%%%%

%%%%%%%%%%%%%%
%\begin{sidewaystable}
\begin{table*}
\centering
\caption{Observations of \Ho\ and \M\ with INTEGRAL and XMM-Newton} 
\label{tab:obs}

\smallskip
\small
\footnotesize

\begin{tabular}{l|l|l|r}
\hline
\multicolumn{1}{c|}{Program} &
\multicolumn{1}{c|}{Revolution or observation} &
\multicolumn{1}{c|}{Dates} &
\multicolumn{1}{c}{Exposure (ks)} \\
\hline
\multicolumn{4}{c}{INTEGRAL} \\
Archive  & 131--133, 179, 180, 250 &
10--17.11.2003,2--6.4, & 750\\   
& & 30--31.10.2004 & \\
0720010 & 856--862, 868-872, 932, 933, 977 & 17.10.--6.12.2009,
1--5.6, & 1930 \\ 
 & & 15--16.10.2010 & \\
0820030 & 1029,1031,1033,1036,1037,1042, & 18.3.--24.5.,
23.9.--2.12.2011 & 1540 \\
& 1046,1048,1049,1051,1092,1093,& \\ 
& 1111,1112,1114,1115 &  & \\  
0920014 & 1225,1226,1228-1231,1233,1234, & 26.10.2012--20.1.2013 & 1690 \\ 
& 1237--1241,1244,1254 & & \\ 
\multicolumn{4}{c}{XMM-Newton} \\
065780 (\Ho) & 2001 & 24.03.2011 & 28 \\
                  & 1601 & 17.04.2011 & 21 \\
                  & 1801 & 26.09.2011 & 25 \\
                  & 2201 & 23.11.2011 & 24 \\
065780 (\M)  & 0101 & 18.03.2011 & 27 \\
                  & 1701 & \,\,\,9.04.2011 & 24 \\
                  & 1901 & 29.04.2011 & 28 \\
                  & 2101 & 24.09.2011 & 23 \\
                  & 2301 & 21.11.2011 & 24 \\
\hline
\end{tabular}
%\end{sidewaystable}
\end{table*}
%%%%%%%%%%%%

Ultraluminous X-ray sources (ULXs) are point-like X-ray sources
observed in extra-nuclear regions of nearby galaxies, whose
luminosities exceed $\sim 2\times 10^{39}$~erg/s. Their nature remains
unknown. Two scenarios are being actively discussed: 
subcritical (at a rate below the Eddington limit) accretion onto a
black hole of intermediate, $(10^2-10^4)\Msun$, mass (IMBH, see,
e.g., \citealt{miletal03}) and supercritical accretion onto a black
hole of stellar, less than a few tens of $\Msun$, mass (StMBH, see,
e.g., \citealt{pouetal07}), possibly with substantial collimation of
radiation toward the observer (e.g., \citealt{king09}). Both
scenarios are of great interest, since in the former case, there
appears an opportunity to explore the conditions that may reflect
an intermediate phase of supermassive black hole growth, and in the
latter case, an extreme regime of gas accretion onto black holes. It
cannot be excluded that the former scenario is realized in some ULXs,
while the latter in others. 

So far, X-ray observations of ULXs have been carried out almost 
exclusively at energies below 10~keV, and many spectra measured in the
2--10~keV energy band could be described by a simple power law with a
slope $\Gamma\sim 2$ (see, e.g., \citealt{kajpou09}). Below 2~keV, an
additional soft component with a color temperature of several
hundred eV was sometimes detected. It was suggested that this
component could be thermal emission from a geometrically thin, optical
thick accretion disk around an IMBH, analogous to the hotter ($\sim
1$~keV) radiation observed from normal X-ray binaries in their
soft/high state. If so, the lower temperature of the disk could imply
\citep{shasun73} a higher mass of the compact object in ULXs compared
to X-ray binaries, whereas the power-law component in ULX spectra
could be attributed to Comptonized radiation from an optically thin,
hot corona of the accretion disk. 

However, this picture faces certain difficulties. In particular, a
number of ULX spectra obtained in long observations by Chandra and
XMM-Newton are poorly described by a power law and have convex shape
in the 2--10~keV energy band. If one describes such spectra in terms of
blackbody radiation from a thin accretion disk, the temperature near
its inner boundary proves to be 2--3~keV. Such values could be
expected, at high accretion rates, for StMBHs, but certainly not for
IMBHs. However, the measured ULX luminosites are
$10^{39}-10^{40}$~erg/s, which exceeds the Eddington limit for a
compact object of stellar mass. Therefore, an alternate scenario for
ULXs has been actively discussed, namely supercritical accretion of
matter onto a StMBH through a geometrically thick disk with a powerful
gas outflow (see, e.g., \citealt{pouetal07}). In this case, the
unusual X-ray spectral shape at energies above 2~keV is
interpreted as the result of Comptonization of the relatively soft
emission from the accretion disk in an optically thick corona and/or 
outflowing wind \citep{stoetal06,glaetal09,fensor11}. 

Measurements in the hard X-ray range could help advance the
understanding of the nature of ULXs. However, such observations were
impossible until recently, because even the brightest ULXs have X-ray
fluxes of less than 1~mCrab in the 2--10~keV energy band and are often
located in sky regions with high number density of X-ray sources. The
capabilities of the INTEGRAL observatory \citep{winetal03}, namely the
combination of relatively high sensitivity and good angular
resolution, have made it possible to overcome these difficulties for
the first time. However, even in this case very long exposures are 
required. Beginning in late 2009, following our proposal, within the
Russian share of observing time, INTEGRAL has been performing deep
observations at energies above $\sim 20$~keV of a sky region
containing the M81 group of galaxies. The targets of these observations are
the nucleus of the M81 galaxy and two well-known ULXs: \Ho\ (also
known as M81~X-9) and \M. Interestingly, although both of these ULXs
are among the brightest over the whole sky in terms of both flux and
luminosity, they are located within one degree of each other. Up to
now, data for nearly 6~Ms worth of observations have been
accumulated. Within the same scientific program, the XMM-Newton X-ray 
observatory performed a series of observations of \Ho\ and \M\ in
2011. The main goal of the coordinated observations by INTEGRAL and
XMM-Newton was to build the X-ray spectra of the aforementioned ULXs in a 
broad energy range from $\sim 200$~eV up to several tens of keV. In
the present paper, we report the results of these observations.   

In what follows, we adopt the distance to \Ho\ to be 3.6~Mpc, which is
the distance to the M81 galaxy \citep{freetal94}, whose satellite 
is the HoIX dwarf galaxy, and that to \M\ to be 3.5~Mpc (the
distance to the M82 galaxy, \citealt{jacetal09}).

Observations of the M81 field started in the 7th cycle of INTEGRAL
observations (AO-7) and were continued in the 8th and 9th cycles (AO-8
and AO-9). Another series of observations of this region of the sky is
planned to take place in late 2013, within the 10th observational
cycle of the INTEGRAL observatory (AO-10). In Table~\ref{tab:obs},
information about the dates and duration of the observations performed
so far is collected. The total accumulated exposure is $\approx 6$~Ms
(nominal exposure, uncorrected for the decrease in efficiency for
off-axis observations of the sources) for the IBIS/ISGRI
instrument. This time includes $\approx 750$~ks of archival data
obtained in 2003--2004, when \Ho\ and \M\ occasionally
fell into the outer regions of the field of view (7--9~deg off-axis)
of the ISGRI detector. These archival data were taken into account in
our analysis, although they barely increase the detection significance 
of the studied objects.   

%%%%%%%%%%%%%%%%%%%%%
\section{Observations}
\label{s:obs}
%%%%%%%%%%%%%%%%%%%%

In 2011, the XMM-Newton observatory carried out two series of
observations of \Ho\ and \M, consisting of 4 and 5 pointings,
respectively, with a duration of about 25~ks each. Table~\ref{tab:obs}
contains some key information about these observations. These 
observations were performed in time intervals that approximately
coincided with the INTEGRAL observations (of much longer duration)
performed during AO-8, hence one can regard the X-ray and hard X-ray
observations carried out in 2011 as quasi-simultaneous.   

In the present work, we used only the data of the IBIS/ISGRI detector
\citep{ubeetal03} from all the data obtained by INTEGRAL. We have also
analyzed the data from the JEM-X instrument, but this has not provided
any additional strong constraints on the X-ray flux from \Ho\ and
\M\ at energies above 10~keV. We therefore built our analysis upon
comparison of the XMM and ISGRI data. 

%%%%%%%%%%%%%%%%%%%%%%%%%%%%%%%%%%%%%%%%%%%%%%%%%%%%
\section{Data reduction, X-ray images}
\label{s:data}
%%%%%%%%%%%%%%%%%%%%%%%%%%%%%%%%%%%%%%%%%%%%%%%%%%%%

%%%%%%%%%%%%%%%%%%%%%
\subsection{INTEGRAL}
\label{s:integral}

Initial reduction of the ISGRI data consisted of reconstruction of sky 
images of the M81 field in a number of energy bands (20--30, 30--45,
45--68, ... keV) using the standard algorithm \citep{krietal10}
utilized in a number of our previous studies. Calibration of measured
source X-ray fluxes was done using archival data of ISGRI observations
of the Crab Nebula. 

\begin{figure}
\centering
\includegraphics[width=\columnwidth]{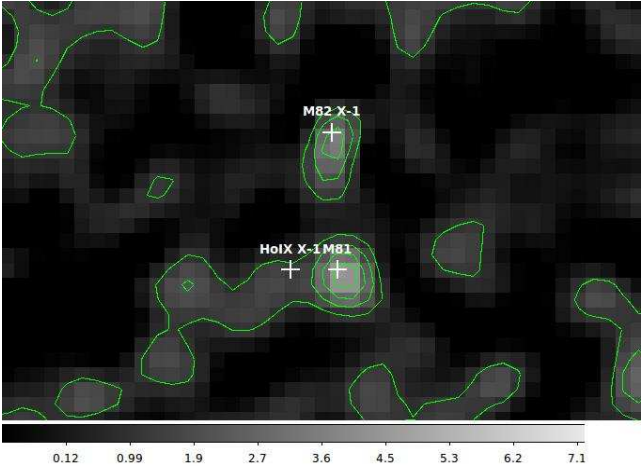}
\caption{Image of the M81 group of galaxies in the 20--30~keV energy band
  obtained by IBIS/ISGRI aboard INTEGRAL over the whole period of
  observations. Plotted is the signal to noise ratio, with the
  contours being 1, 2, 3, and 4$\sigma$. Denoted are the positions of
  \Ho, \M, and the nucleus of the M81 galaxy. The distance between
  \Ho\ and M81 is 13'.}
\label{fig:int_image}
\end{figure}

Figure~\ref{fig:int_image} shows an image of the M81 field in the
20--30~keV energy band, constructed using all the data obtained by
ISGRI from October 2009 to January 2013. The \M\ source is
detected at a $3.6\sigma$ significance level. Its flux in the
20--30~keV energy band is $(0.44\pm0.12)$~mCrab, which corresponds to
$(2.1\pm 0.6)\times 10^{-12}$~erg/s/cm$^2$, assuming a Crab-like
spectrum. Under the assumption of isotropical radiation, the
luminosity of \M\ in the 20--30~keV range is $(3.1\pm0.9)\times
10^{39}$~erg/s. \M\ is practially undetected ($1.6\sigma$) in the next
energy band (30--45~keV), nor in the higher energy bands. The
\Ho\ source is not detected even in the softest ISGRI energy band, the
upper limit (2$\sigma$) on its 20--30~keV flux being 0.25~mCrab, or
$1.2\times 10^{-12}$~erg/s/cm$^2$ (assuming a Crab-like
spectrum). Hence, the luminosity of \Ho\ in the 20--30~keV energy
band does not exceed $1.9\times 10^{39}$~erg/s.
 
At a distance of just 13~arcmin from \Ho, there is a relatively bright
hard X-ray source whose position is consistent with that of the
nucleus of the M81 galaxy. This object belongs to the class of
low-luminosity active galactic nuclei and its study in the hard X-ray
range is of great interest but beyond the scope of the present
work. Because of the proximity to M81, studies of \Ho\ in the hard
X-ray range were practically impossible before 
the advent of INTEGRAL. The angular resolution of the BAT telescope
aboard Swift is also insufficient for this purpose.  

\begin{figure}
\centering
\includegraphics[bb=10 150 560 670,width=\columnwidth]{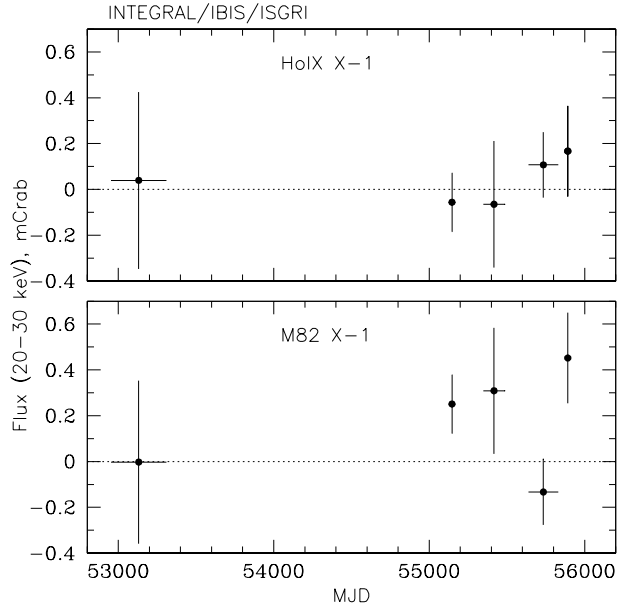}    
\caption{Light curves of \Ho\ and \M\ in the 20--30~keV energy band
  over the whole period of INTEGRAL observations.} 
\label{fig:int_lcurve}
\end{figure}

Figure~\ref{fig:int_lcurve} shows the light curves of \Ho\ and \M\ in
the 20--30~keV energy band obtained using the ISGRI data. It is seen
that the sources remained weaker than $\sim 0.7$~mCrab during the 
observations in 2003--2013, although the data are of insufficient
quality to judge about the variability of the studied objects. 

\subsection{XMM-Newton}
\label{s:xmm}

The reduction of the data obtained by the MOS1, MOS2, and pn
instruments aboard XMM-Newton was done using the standard software
SAS v.11\footnote{http://xmm2.esac.esa.int/sas/}. Since thee were no strong
proton flares during our observations, data filtering was done
in a standard way.  

\begin{figure}
\centering
\includegraphics[width=0.85\columnwidth]{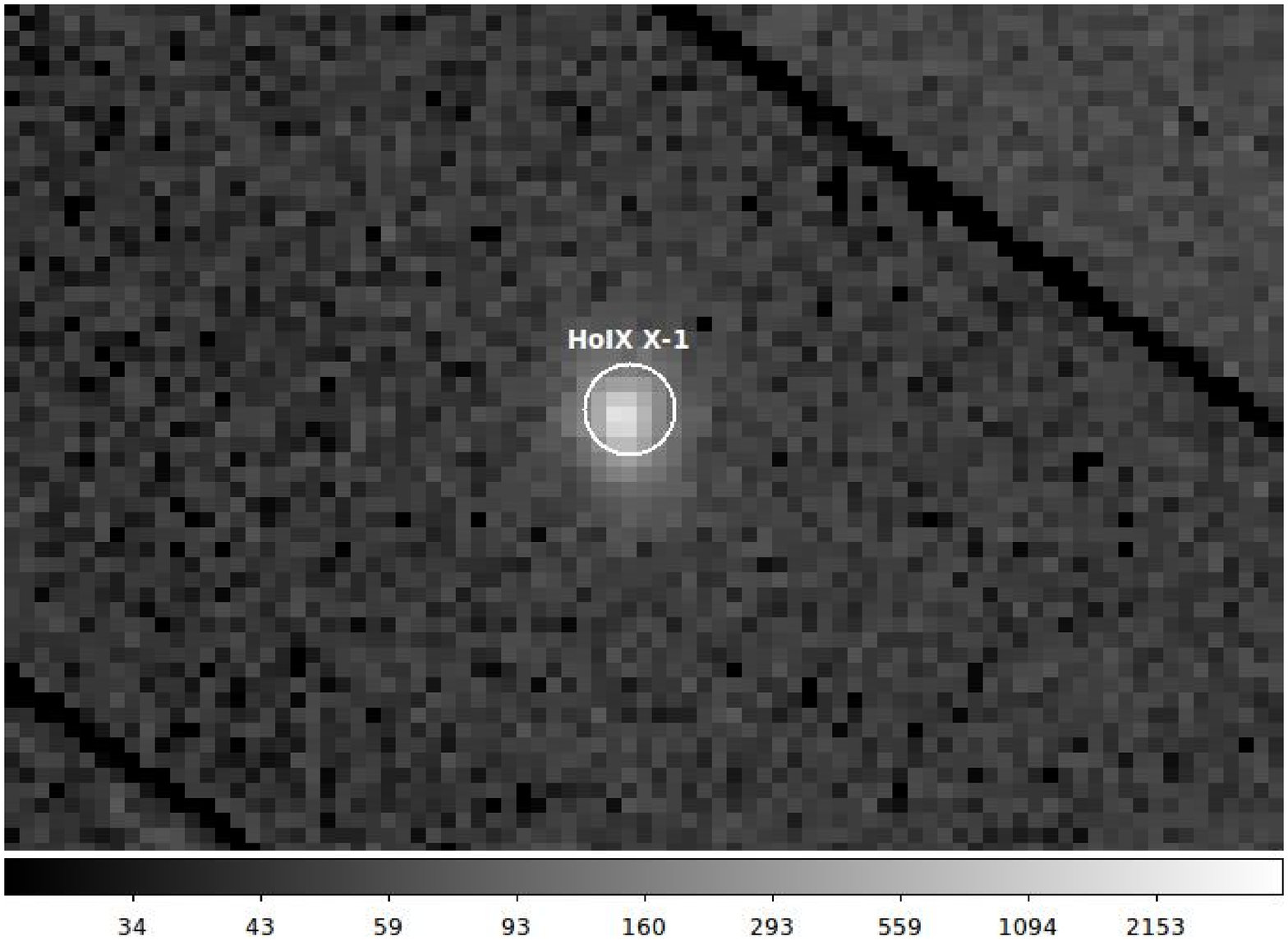}
\includegraphics[width=0.85\columnwidth]{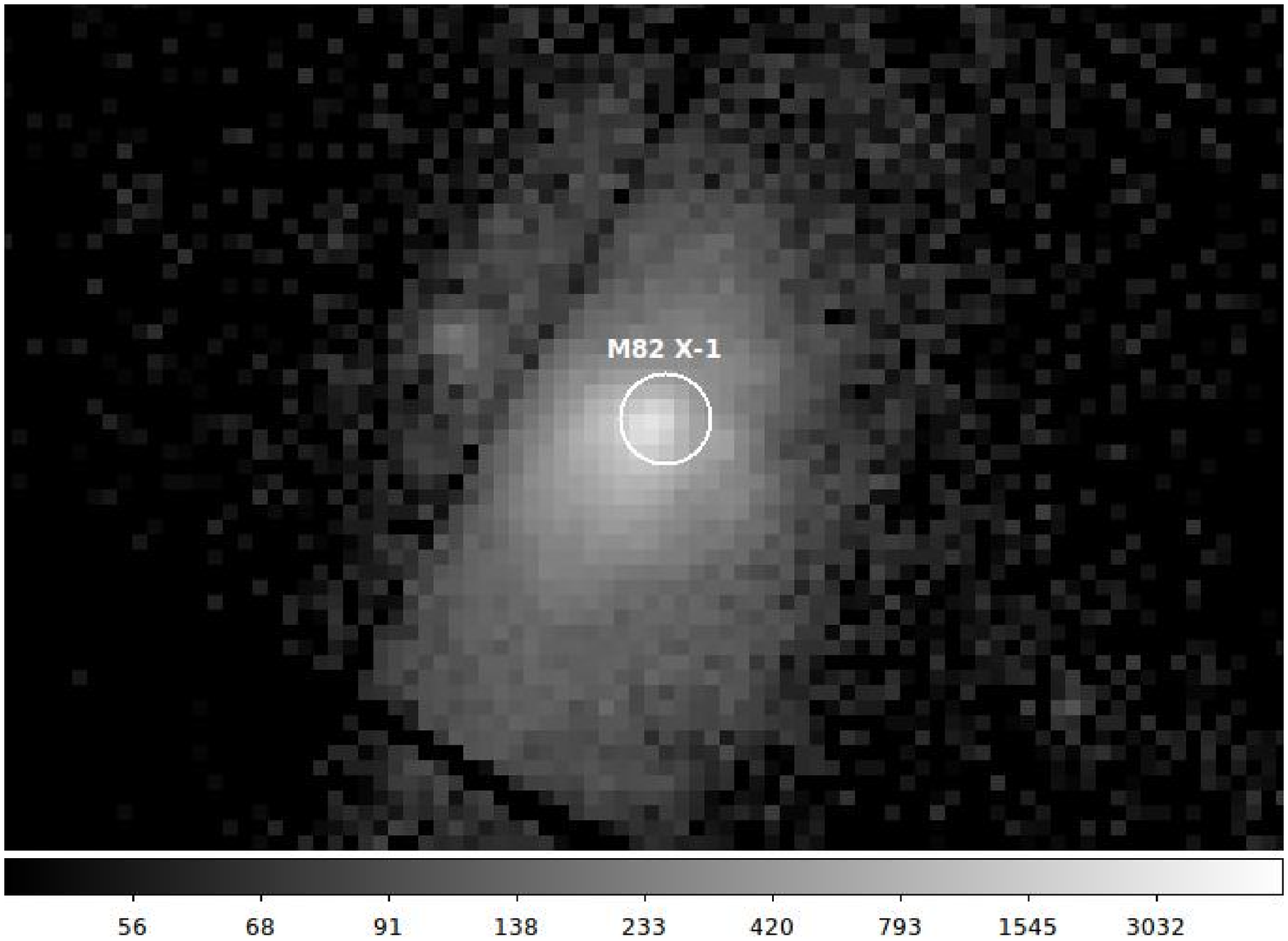}
\caption{X-ray images of the sky (shown are counts in the 0.2--13~keV
  energy band) near \Ho\ and \M\ obtained using data of the pn
  detector aboard XMM-Newton in individual observations (on March 24
  and 18, 2011, respectively). The spectral analysis was based on the
  counts recorded within the 13''-radius circles shown in the
  figure.}
\label{fig:xmm_image}
\end{figure}

Figure~\ref{fig:xmm_image} shows examples of images obtained with the
pn detector of the EPIC camera in individual observations of \Ho\ and
\M\ in 2011. A bright X-ray source is detected in the direction of
\Ho, whose angular size is compatable with the signal expected from a
point source. There are no other noticeable X-ray sources near
\Ho. This probably reflects the fact that the object is located on the
outskirts of the dwarf galaxy HoIX (see, e.g.,
\citealt{grietal11}). Similarly, there 
is strong emission from the direction of \M, which can be attributed
to the ULX at hand, however this signal is overlaid on a bright
background created by the M82 galaxy. This background likely
consists of a large number of unresolved faint objects and diffuse
emission from the interstellar gas. As is shown below, the contribution
of the galactic background can be taken into account during the
analysis of the \M\ X-ray spectrum obtained from the XMM-Newton data.   
 
For the subsequent spectral analysis, we used the counts recorded
within the circles of 13'' radius around \Ho\ and \M\ (see
Fig.~\ref{fig:xmm_image}). This region contains $\sim 70$\% of
all the photons from a point source, and this apperture coefficient
was taken into account during the spectral analysis. To estimate the
background unrelated to the emission of the host galaxy, we used
regions located at distances $\sim 1$--3' from the studied
objects. Because \M\ is located in the central region of the M82
galaxy, even the compact (13'') spectrum-extraction region contains a
number of other fairly bright X-ray sources, known from
arcsec-resolution observations with the Chandra
telescope. Unfortunately, it is impossible to remove the contribution
of these sources during the analysis of the XMM-Newton data. However,
the Chandra data suggest that this contribution is unlikely to exceed
$\sim 30$\% \citep{matetal01}. 

%%%%%%%%%%%%%%%%%%%%%%%%%%%%%
\section{Spectral analysis}
\label{s:spectra}
%%%%%%%%%%%%%%%%%%%%%%%%%%%%%

%%%%%%%%%%%%%%%%
\subsection{\Ho}

\begin{figure}
\centering
\includegraphics[bb=0 160 570 670,width=\columnwidth]{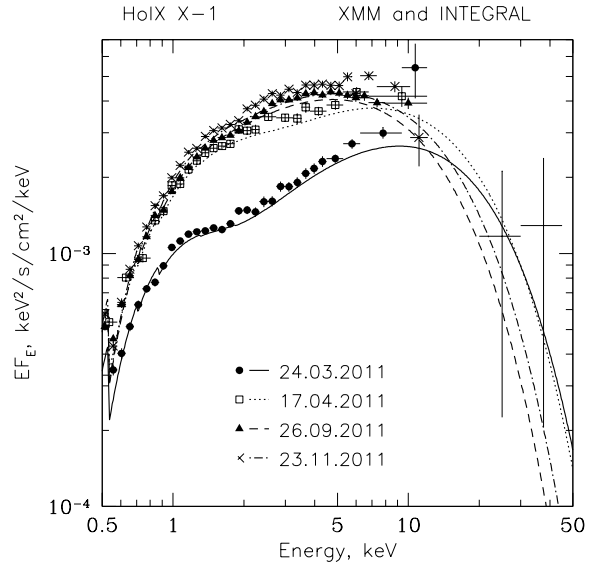}
\caption{X-ray spectra of \Ho\ measured by XMM-Newton during four
  observations in 2011 and hard X-ray fluxes in the 20--35 and
  35--50~keV energy bands obtained from ISGRI data averaged over the whole
  period of observations. Different curves show the results of joint
  fitting of the XMM-Newton and ISGRI data by the
  \textit{wabs(diskbb+cutoffpl)} model in XSPEC.}
\label{fig:ho_spec}
\end{figure}

Figure~\ref{fig:ho_spec} shows the X-ray spectra of \Ho\ measured with
XMM-Newton during the four observations carried out in 2011 (see
Table~\ref{tab:obs}), as well as the hard X-ray fluxes in the 20--35
and 35--50~keV energy bands obtained using the ISGRI data for the
whole period of INTEGRAL observations.

We first carried out a spectral analysis of the XMM-Newton data only, 
in the 0.2--13~keV energy band. To this end, we used the XSPEC package
\citep{arnaud96}. A given model was fitted simultaneously to the data
from all three X-ray cameras, pn, MOS1, and MOS2, allowing the model
normalization to differ between the detectors. The resulting
differences in the normalization proved to be less than 10\%, and the
data of the three detectors showed a good mutual agreement. Each of the
four spectra shown in Fig.~\ref{fig:ho_spec} is the result of
averaging over the pn, MOS1, and MOS2 data; a similar averaging was done 
for the corresponding spectral models. 

As can be seen from Fig.~\ref{fig:ho_spec}, the X-ray spectrum of
\Ho\ varied significantly over 2011. The first spectrum, obtained 
on March 24, exhibits a local maximum near 1~keV and a power-law
continuum at higher energies. This spectrum is well fit (see
Table~\ref{tab:ho_fits}) by a sum of a model of blackbody radiation
from a multi-temperature accretion disk and a power law with an
exponential cutoff at high energies, modified by absorption along the
line of sight, i.e., by the \textit{wabs(diskbb+cutoffpl)} model in
XSPEC. However, the statistical significance of the detection of a
cutoff is low: replacement of the \textit{powerlaw} model by
\textit{cutoffpl}, i.e., addition of one degree of freedom, leads to a
decrease of the $\chi^2$ value by only 6. The paramaters of the
\textit{diskbb} model are a normalization and the temperature ($kT_{\rm
  in}$) at the inner boundary of the multicolor disk. Our analysis
assumed that the disk is observed along its axis. 

%%%%%%%%%%%%%%
%\begin{sidewaystable}
\begin{table*}
\centering
\caption{Results of spectral analysis for \Ho}
\label{tab:ho_fits}

\smallskip
\footnotesize

\begin{tabular}{l|c|c|c|c}
\hline
Parameter & XMM2001 (+ISGRI) & XMM1601 (+ISGRI) & XMM1801 (+ISGRI) &
XMM2201 (+ISGRI) \\ 
\hline
\multicolumn{5}{c}{wabs(diskbb+powerlaw)}\\
$\Gamma$ & $1.47\pm0.04$ ($1.55\pm0.03$) & $1.81\pm0.09$ ($1.89\pm0.03$) &
$1.871\pm0.017$ ($1.872\pm0.010$) & $1.863\pm0.019$ ($1.921\pm0.019$) \\   

$\chi^2$ (d.o.f) & 947.1/936 (979.9/938) & 642.3/626
(664.6/628) & 1580.9/1464 (1570.0/1466) & 1281.5/1273 
(1514.0/1275) \\   
\multicolumn{5}{c}{wabs(diskbb+cutoffpl)}\\
$\NH$, $10^{22}$~cm$^{-2}$ & $0.179\pm0.015$ & $0.15\pm0.03$ &
$0.193\pm0.004$ & $0.204\pm0.005$ \\  

$F_{\rm bb}$ (0.2--10~keV), & & & \\
$10^{-12}$~erg/s/cm$^2$ & $2.59\pm0.15$ & $2.8\pm1.2$ & $1.4\pm0.4$ &
$1.7\pm0.5$ \\ 

$L_{\rm bb}$ (0.2--10~keV), & & & \\
$10^{39}$~erg/s & $4.0\pm0.2$ & $4.3\pm1.9$ & $2.2\pm0.6$ &
$2.7\pm0.7$ \\

$kT_{\rm in}$, keV & $0.29\pm0.03$ & $0.48\pm0.06$ & 0.3 (fixed) & 0.3
(fixed) \\ 

$F_{\rm pl}$ (0.2--10~keV), & & & \\
$10^{-12}$~erg/s/cm$^2$ & $7.5\pm0.3$ & $12.9\pm1.9$ & $16.9\pm0.3$ &
$18.2\pm0.4$ \\

$L_{\rm pl}$ (0.2--10~keV), & & & \\
$10^{39}$~erg/s & $11.6\pm0.5$ & $20\pm3$ & $26.2\pm0.5$ &
$28.3\pm0.6$ \\

$\Gamma$ & $1.0\pm0.2$ ($1.01\pm0.17$) & $1.1\pm0.6$ ($1.2\pm0.3$) &
$1.28\pm0.08$ ($1.29\pm0.08$) & $1.33\pm0.10$ ($1.35\pm0.09$) \\  

$E_{\rm cut}$, keV & $9\pm4$ ($9\pm3$) & $8\pm7$ ($9\pm4$) &
$6.4\pm0.9$ ($6.5\pm0.9$) & $7.2\pm1.3$ ($7.4\pm1.3$) \\  

$\chi^2$ (d.o.f.) & 941.1/935 (941.7/937) & 641.3/625
(641.9/627) & 1454.8/1463 (1456.1/1465) & 1242.7/1272
(1243.7/1274) \\ 
\multicolumn{5}{c}{wabs(diskbb+compst)}\\
$kT$, keV & $2.6\pm0.3$ ($2.7\pm0.4$) & $5\pm4$ ($3.3\pm0.8$) &
$2.25\pm0.12$ ($2.31\pm0.13$) & $2.31\pm0.16$ ($2.33\pm0.16$) \\   

$\tau$ & $16.9\pm 1.6$ ($16.4\pm1.5$) & $10\pm5$ ($13\pm2$) &
$14.7\pm0.6$ ($14.4\pm0.6$) & $14.5\pm0.7$ ($14.5\pm0.7$) \\  

$\chi^2$ (d.o.f.) & 944.4/935 (946.2/937) & 641.5/625
(642.6/627) & 1473.5/1463 (1478.2/1465) & 1248.1/1272
(1250.3/1274) \\  
\hline
\end{tabular}

%\end{sidewaystable}
\end{table*}
%%%%%%%%%%%%

In the subsequent XMM-Newton observation (April 17, 2011), the
blackbody component is barely detected (the detection
significance is just above 2$\sigma$), nor is detected a high-energy
cutoff of the power-law continuum. Hence, this spectrum can be
described, over the 0.2--13~keV energy band, by a simple power law with
absorption. In the latest two observations (September 26 and November
23, 2011), the soft blackbody component is again barely detected (the
detection significance is $\sim 3$--$4\sigma$ when the 
temperature at the inner boundary of the disk is fixed at 0.3~keV),
however a rollover of the power-law continuum becomes evident above
$\sim 5$--10~keV  (see Fig.~\ref{fig:ho_spec} and
Table~\ref{tab:ho_fits}).

Table~\ref{tab:ho_fits} summarizes, for all four XMM-Newton
observations, the unabsorbed fluxes and luminosities of the blackbody
and power-law (with a cutoff) spectral components in the 0.2--10~keV
energy band ($F_{\rm bb}$, $F_{\rm pl}$, $L_{\rm bb}$, $L_{\rm
  pl}$). The total luminosity of \Ho\ in this energy range increased
over the observing campaign from $\sim 1.5\times 10^{40}$ to $\sim 3\times
10^{40}$~erg/s, the luminosity of the blackbody component being $\sim
(2-4)\times 10^{39}$~erg/s. The deduced line-of-sight absorption
columns, $\NH\sim (1.5-2)\times 10^{21}$~cm$^{-2}$, somewhat exceed
the corresponding value for the interstellar absorption through the
Galaxy in the direction of the M81 group of galaxies, $\sim 5\times
10^{20}$~cm$^{-2}$ \citep{kaletal05}, indicating the presence of cold
gas near \Ho. 

The presence of a strong cutoff in the spectrum of \Ho\ at energies
above $\sim 5$--10~keV becomes evident from comparison of the
XMM-Newton data with the upper limits 
on the hard X-ray flux in the 20--30 and 30--45~keV energy bands
obtained with the ISGRI detector aboard INTEGRAL (see
Fig.~\ref{fig:ho_spec}). Joint fitting of the data of both
observatories allows us to fully exclude a
power-law spectral model from consideration and reliably establish
the position of the 
cutoff in the hard X-ray spectrum: $E_{\rm cut}\sim 8$~keV (see
Table~\ref{tab:ho_fits}). Reasonable values of $\chi^2$ per degree of
freedom obtain when the ISGRI data are fitted jointly with any of the
four  XMM-Newton spectra. However, one should keep in mind that the
latter data, obtained in individual observations 
lasting less than a day, are here compared with ISGRI measurements
averaged over several Ms and accumulated in various years.

A natural mechanism to explain power-law spectra with a high-energy
cutoff is Comptonization of soft radiation on hot electrons. We thus
tried to describe the XMM-Newton spectra of \Ho\ by a model that,
together with a blackbody component, includes a component of
Comptionized radiation emergent from a cloud of hot gas
\citep{suntit80} -- the \textit{wabs(diskbb+compst)} model in
XSPEC. This model desribes the data well, albeit somewhat worse that the
\textit{wabs(diskbb+cutoffpl)} model (see
Table~\ref{tab:ho_fits}). The resulting gas temperature is $\sim
2$--3~keV and the optical depth of the cloud is $\sim 15$. 

%%%%%%%%%%%%%%%
\subsection{\M}

\begin{figure}
\centering
\includegraphics[bb=0 160 570 670,width=\columnwidth]{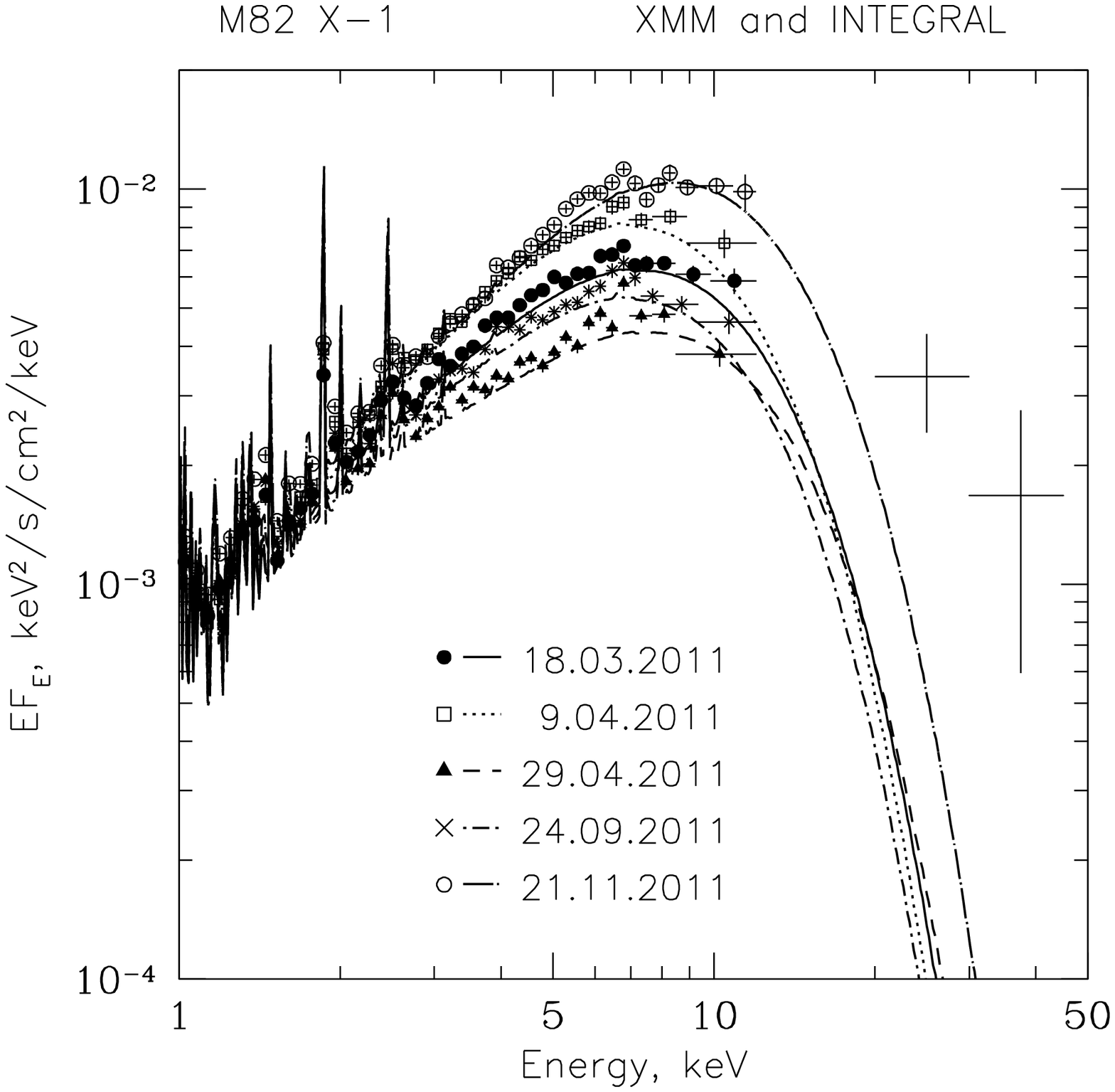}   
\caption{X-ray spectra of \M\ measured by XMM-Newton during five
  observations in 2011 and fluxes in the 20--35 and 35--50~keV energy
  bands obatined from ISGRI data averaged over the whole period of
  observations. Different curves show the results of fitting of the
  XMM-Newton data by the \textit{wabs(diskbb+compst)} model in XSPEC.
}
\label{fig:m_spec}
\end{figure}

%%%%%%%%%%%%%%
%\begin{sidewaystable}
\begin{table*}
\centering
\caption{Results of spectral fitting for \M}
\label{tab:m_fits}

\smallskip
\footnotesize

\begin{tabular}{l|c|c|c|c|c}
\hline
Parameter & XMM0101 & XMM1701 & XMM1901 & XMM2101 & XMM2301 \\   
 & (+ISGRI) & (+ISGRI) & (+ISGRI) & (+ISGRI) & (+ISGRI) \\   
\multicolumn{6}{c}{wabs(apec+cutoffpl)} \\
$\Gamma$ & $0.21\pm0.09$ & $-0.15\pm0.12$ & $0.58\pm0.12$ &
$-0.05\pm0.13$ & $-0.20\pm0.10$ \\  
 & ($0.29\pm0.08$) & ($-0.04\pm0.11$) & ($0.74\pm0.10$) &
($0.04\pm0.12$) & ($-0.13\pm0.08$) \\   

$E_{\rm cut}$, keV & $4.4\pm0.4$ & $3.4\pm0.3$ & $6.2\pm0.9$ &
$3.5\pm0.3$ & $4.1\pm0.3$ \\ 
 & ($4.8\pm0.4$) & ($3.7\pm0.3$) & ($7.7\pm1.1$) & ($3.7\pm0.3$) &
($4.3\pm0.3$) \\  

$\chi^2$ (d.o.f.) & 1547.9/1491 & 1315.4/1326 & 1433.2/1270 &
1294.8/1219 & 1746.2/1657 \\ 
 & (1554.3/1493) & (1323.5/1328) & (1436.3/1272) & (1305.4/1221) &
(1748.0/1659) \\
\multicolumn{6}{c}{wabs(apec+compst)}\\
$\NH$, $10^{22}$~cm$^{-2}$ & $1.15\pm0.03$ & $1.22\pm0.04$ &
$1.20\pm0.02$ & $1.25\pm0.03$ & $1.19\pm0.03$ \\ 

$F_{\rm apec}$ (1--10~keV), & & & & & \\ 
$10^{-12}$~erg/s/cm$^2$ & $5.2\pm0.2$ & $5.8\pm0.3$ & $6.7\pm0.2$ &
$7.1\pm0.3$ & $7.0\pm0.3$ \\ 

$L_{\rm apec}$ (1--10~keV), & & & & & \\
$10^{39}$~erg/s & $7.6\pm0.3$ & $8.5\pm0.4$ & $9.8\pm0.3$ &
$10.4\pm0.4$ & $10.3\pm0.4$ \\

$kT_{\rm apec}$, keV & $0.880\pm0.015$ & $0.923\pm0.019$ &
$0.922\pm0.013$ & $0.947\pm0.014$ & $0.910\pm0.014$ \\ 

$F_{\rm comp}$ (1--10~keV), & $13.93\pm0.12$ & $17.41\pm0.17$ &
$10.08\pm0.12$ & $11.90\pm0.13$ & $19.80\pm0.15$ \\

$L_{\rm comp}$ (1--10~keV), & $20.42\pm0.18$ & $25.5\pm0.3$ &
$14.78\pm0.17$ & $17.44\pm0.19$ & $29.0\pm0.2$ \\

$kT$, keV & $2.11\pm0.05$ & $1.94\pm0.05$ & $2.29\pm0.09$ &
$2.00\pm0.06$ & $2.33\pm0.05$ \\    
 & ($2.12\pm0.05$) & ($1.96\pm0.05$) & ($2.32\pm0.10$) &
($2.01\pm0.06$) & ($2.36\pm0.05$) \\   

$\tau$ & $24.2\pm0.8$ & $28.5\pm1.3$ & $21.0\pm1.0$ & $25.5\pm1.2$ &
$26.2\pm0.9$ \\     
 & ($24.1\pm0.8$) & ($28.2\pm1.3$) & ($20.7\pm1.0$) & ($25.3\pm1.2$) &
($25.9\pm0.9$) \\ 

$\chi^2$ (d.o.f.) & 1506.9/1491 & 1297.6/1326 & 1389.69/1270 &
1275.7/1219 & 1654.5/1657 \\     
 & (1518.5/1493) & (1309.60/1328) & (1401.1/1272) & (1288.6/1221) &
(1660.2/1659) \\ 
\multicolumn{6}{c}{wabs(apec+diskbb)}\\
$kT_{\rm in}$, keV & $3.41\pm0.07$ & $3.59\pm0.10$ & $3.36\pm0.10$ &
$3.23\pm0.08$ & $4.65\pm0.12$ \\ 
 & ($3.43\pm0.07$) & ($3.63\pm0.10$) & ($3.39\pm0.10$) &
($3.25\pm0.08$) & ($4.56\pm0.10$) \\  

$\chi^2$ (d.o.f.) & 1541.5/1492 & $1336.6/1327$ & 1425.2/1271 &
1298.8/1220 & 1763.8/1658 \\  
 & (1549.3/1494) & (1340.7/1329) & (1435.2/1273) & (1308.6/1222) &
(1767.9/1660) \\  

\hline
\end{tabular}

%\end{sidewaystable}
\end{table*}
%%%%%%%%%%%%

We carried out a similar analysis for \M. There is a significant
difference with respect to the previous case, namely that the spectra
obtained with XMM-Newton contain an unsubtracted contribution from the
background emission associated with the host galaxy M82. The presence
of a large number of strong emission lines clearly suggests that this
additional emission is thermal radiation from an optically thin, hot
plasma. We have thus added an APEC \citep{smietal01} component into
the spectral model, assuming Solar chemical composition and further
restricted our consideration by energies above 
1~keV, since at lower energies the contribution of the background
emission becomes dominat and does not enable reliable extraction of
the emission associated with \M. This, unfortunately, precludes
finding out whether the \M\ spectrum contains a soft blackbody
component similar to that found in \Ho. 

To describe all five XMM-Newton spectra of \M, it is necessary to take
into account photoabsorption at low energies, with $\NH\sim 1.2\times
10^{22}$~cm$^{-2}$. This value significantly exceeds the interstellar
absorption through the Galaxy in the given direction, $\sim 5\times 
10^{20}$~cm$^{-2}$ \citep{kaletal05}, indicating the presence of cold
gas near \M. All five spectra clearly exhibit a rollover at high
energies, hence we used the following spectral model in our analysis:
\textit{wabs(apec+cutoffpl)}. The results of fitting by this model
prove to be quite satisfactory (based on $\chi^2$) and are quoted in
Table~\ref{tab:m_fits}. Even better results obtain when using a model
of Comptonized radiation from a hot cloud \citep{suntit80}, i.e., when
fitting the spectra by the \textit{wabs(apec+compst)} model in XSPEC.  

We have also attempted to describe the hard X-ray continuum by an
alternate model of blackbody radiation from a hot ($kT_{\rm in}\sim
3$--4~keV) accretion disk, i.e, to apply the
\textit{wabs(apec+diskbb)} model. This model describes some of the
spectra nearly as well as the \textit{wabs(apec+cutoffpl)} model,
albeit significantly worse than the \textit{wabs(apec+compst)} model. 

The addition of ISGRI spectral points at energies above 20~keV leads
to an interesting result. All the model spectra determined by fitting
the XMM-Newton data lie below the ISGRI points. This can be
inetrpreted as the presence of an additional hard X-ray component in
the source's spectrum. However, the statistical significance of this
excess is low (about $3\sigma$). It is not unlikely that the
additional hard X-ray flux is due to other point sources in the
central region of the M82 galaxy, which cannot be separated from
\M\ using XMM-Newton data. In any case, comparison of the hard X-ray
flux measured by ISGRI with the XMM-Newton spectra clearly implies a
cutoff in the X-ray spectrum of \M\ at energies above $\sim 10$~keV.
 
%%%%%%%%%%%%%%%%%%%%
\section{Discussion}
\label{s:discuss}
%%%%%%%%%%%%%%%%%%%%

Using data of the XMM-Newton and INTEGRAL observatories, we have
detected a cutoff at energies above 10~keV in the spectra of the
ultraluminous X-ray sources \Ho\ and \M. Previously, there were
indications of the existence of such a cutoff in the spectra of these
sources from observations with the Suzaku observatory
\citep{miyetal09,dewetal13}, however the crude angular resolution of
the PIN instrument precluded reliable separation of the hard X-ray
fluxes of the ULXs and the low-luminosity active galactic nucleus M81.  

The derived broad-band X-ray spectra can be well described by a model
of Comptonization of radiation in a cloud of gas of moderate
temperature and large optical depth: $kT\sim 2.5$~keV, $\tau\sim
15$ for \Ho, and $kT\sim 2$~keV, $\tau\sim 25$ for \M. These values
indicate that the Comptonization takes place in a nearly saturated 
regime (the Comptonization parameter $y=(4kT/m_ec^2)\tau^2\sim
4$--10, see, \citealt{suntit80}). Such conditions are quite unusual for
the hot coronae of accretion disks in normal X-ray binaries (where
accretion onto a StMBH at a subcritical rate occurs), but can
be fulfilled during supercritical accretion of matter onto an IMBH
accompanied by a strong outflow of gas from the central regions
\citep{shasun73,ohsetal05,pouetal07}. In this case, the soft emission
component with a characteristic temperature of $kT_{\rm in}\sim
0.3$~keV detected from \Ho\ by XMM-Newton can be related to 
regions of a thin accretion disk that are open to the observer (beyond
the spherization radius of the accretion flow) or with the photosphere
of the wind outflowing from the disk. 

Therefore, the reported results of XMM-Newton and INTEGRAL
observations confirm the existence of a spectral state
\citep{stoetal06,glaetal09} specific to ultraluminous X-ray 
sources, which is unlike any of the spectral states known for normal
X-ray binaries. 

For \M, we have found an indication of an additional hard
X-ray emission component. This radiation may hint at the presence in
the inner part of the supercritical accretion 
flow of a hot corona similar to the coronae of accretion disks in
X-ray binaries. However, this excess may also be associated with the
unsubtracted contribution of other fairly bright X-ray sources in
the central region of the starforming galaxy M82, which are resolved
by Chandra \citep{matetal01}. 

Given the findings of this study, we are awaiting with great
interest the upcoming announcement of the results of hard X-ray
observations of a number of ULXs by the new-generation observatory
NuSTAR, whose high sensitivity and angular resolution enable detailed
studies of ULX spectral properties at energies above 10~keV for the
first time. 

\begin{acknowledgements}
This research was partially supported by the Ministry of Science and
Education of the Russian Federation (contract 8701), Russian
Foundation for Basic Research (grant 13-02-01365), programs P-21 and
OFN-17 of the Presidium of the Russian Academy of Sciences, and program
NSh-6137.2014.2 for support of leading scientific schools in
Russia. The research made use of data obtained from the Russian
INTEGRAL Data Center and XMM-Newton Data Center. The authors are
grateful to Eugene Churazov for developing the methods of analysis of
INTEGRAL/IBIS data and providing the software. 

\end{acknowledgements}

\end{document}